\documentclass[conference]{IEEEtran}

\usepackage{amssymb,amsmath,amsfonts,bm,epsfig,graphicx,theorem,epstopdf,wasysym}
\usepackage{rotating,setspace,latexsym,epsf,color,dsfont,caption}
\usepackage{cite,subfigure}
\usepackage{algorithm,algpseudocode}

\newtheorem{Theorem}{Theorem}
\newtheorem{Corollary}{Corollary}

\newtheorem{Proposition}{Proposition}
\newtheorem{Definition}{Definition}

\newenvironment{Proof}[1]{\medskip\par\noindent
{\bf Proof:\,}\,#1}{{\mbox{\,$\blacksquare$}\par}}

\makeatletter
\newcommand\fs@spaceruled{\def\@fs@cfont{\bfseries}\let\@fs@capt\floatc@ruled
  \def\@fs@pre{\vspace{5\baselineskip}\hrule height.8pt depth0pt \kern2pt}%
  \def\@fs@post{\kern2pt\hrule\relax}%
  \def\@fs@mid{\kern2pt\hrule\kern2pt}%
  \let\@fs@iftopcapt\iftrue}
\makeatother

\newcommand{\Eb}{\mathbb{E}}
\newcommand{\Pb}{\mathbb{P}}
\newcommand{\Zb}{\mathbb{Z}}
\newcommand{\Nb}{{\mathbb N}}

\newcommand{\lv}{\mathbf{1}}

\newcommand{\Cc}{\mathcal{C}}
\newcommand{\Xc}{\mathcal{X}}

\newcommand{\tcr}{\textcolor{red}}

\title{Information Freshness for Timely Detection of Status Changes}
\author{
	\IEEEauthorblockN{Songtao Feng \qquad Jing Yang}
	\IEEEauthorblockA{School of Electrical Engineering and Computer Science \\
		The Pennsylvania State University\\
		University Park, PA 16802\\
		\emph{\{sxf302,yangjing\}@psu.edu}}

	\thanks{This work was supported in part by the US National Science Foundation (NSF) under Grant ECCS-1650299.}
}

\begin{document}
\IEEEoverridecommandlockouts
\maketitle
\thispagestyle{empty}

\begin{abstract}
In this paper, we aim to establish the connection between Age of Information (AoI) in network theory, information uncertainty in information theory, and detection delay in time series analysis. We consider a dynamic system whose state changes at discrete time points, and a state change won't be detected until an update generated after the change point is delivered to the destination for the first time. We introduce an information theoretic metric to measure the information freshness at the destination, and name it as generalized Age of Information (GAoI). We show that under any state-independent online updating policy, if the underlying state of the system evolves according to a stationary Markov chain, the GAoI is proportional to the AoI. Besides, the accumulative GAoI and AoI are proportional to the expected accumulative detection delay of all changes points over a period of time. Thus, any (G)AoI-optimal state-independent updating policy equivalently minimizes the corresponding expected change point detection delay, which validates the fundamental role of (G)AoI in real-time status monitoring. Besides, we also investigate a Bayesian change point detection scenario where the underlying state evolution is not stationary. Although AoI is no longer related to detection delay explicitly, we show that the accumulative GAoI is still an affine function of the expected detection delay, which indicates the versatility of GAoI in capturing information freshness in dynamic systems.

\end{abstract}


\section{Introduction}
Recently, how to quantify and optimize information freshness in sensing, communication and computing systems has attracted growing attention from different communities. Various metrics have been proposed to measure information freshness. Among them, perhaps the most prevalent one is \textit{Age of Information} (AoI)~\cite{infocom/KaulYG12}. 
Originally proposed for a single-node monitoring system that keeps sending time-stamped status updates to a destination, the metric quantifies the freshness of information from the destination's perspective. 
Specifically, at time $t$, the AoI at the destination is defined as $t-\delta(t)$, where $\delta(t)$ is the generation time (time stamp) of the freshest received update at the destination. 
With this definition, the AoI {performance has} been characterized for various updating policies in different systems ~\cite{YatesK16,Pappas:2015:ICC,isit/NajmN16,tit/KamKNE16}. AoI optimization from the perspectives of sampling, transmission scheduling and coding has also been investigated under various system constraints  ~\cite{Sun:ISIT:2017,Jiang:2019:INFOCOM,isit/BedewySS16,infocom/SunUYKS16,He:2018:TIT,Kadota:2018:INFOCOM,Wang:JCN:2019,Zhou:TWC:2019,Yang:2017:AoI,Yang:2018:IT,Feng:TAC:arXiv,Yates:2017:ISIT,Mayekar2018,Emina:2018,Feng:Globecom:2019,Baknina:2018:CISS}.

\if{0}
In recent years, several work for evaluating information freshness about time-correlated sources have been discussed in~\cite{isit/YatesK12,SunPU17,infocom/SunUYKS16,Kosta:2017:nonlinear,isit/BedewySS16,shroff_age_multi_hop}. \tcr{add more, check Sun Yin's paper}

The most popular information freshness metric AoI, though characterizes the timeliness of the information at the destination, it excludes the information carried by the underlying system which helps determine the system status in many cases. 
Reference \cite{Sun:2018:SPAWC} proposes to use the mutual information between the real-time source value and the delivered samples at the receiver to quantify the freshness of the information contained in the delivered samples. It shows that for a time-homogeneous Markov chain, the mutual information can be expressed as a non-negative and non-increasing function of the age.
In~\cite{singh2019optimal}, a metric called ``Value of Information" (VoI) is introduced to facilitate packet scheduling for low-error Kalman filter based estimation. VoI depends on the age of the packet, as well as the mutual information between the packet content and the system status, which is equivalent to the variance of the noise associated with the measurement. 
Generally speaking, it lacks a systematic way to relate information freshness with the dynamics of the underlying monitored signal.
\fi

Even though AoI can be used as a measure of ``staleness" of the status information in a system, it does not take the dynamics of the underlying status evolution into consideration. 
Recently, there are a few attempts to take the properties of the monitored signal into the definition of information freshness. 
A metric called ``{Age of Synchronization}'' (AoS) is proposed in \cite{Zhong:ISIT:2018}. 
It refers to the duration since the destination became \textit{desynchronized} with the source.
In the same spirit, another metric called ``Age of Incorrect Information'' (AoII) is proposed in \cite{maatouk2019age}.
AoII takes both the time that the monitor is unaware of the correct status of the system and the difference between the current estimate at the monitor and the actual state of system into the definition. With particular penalty functions, AoII reduces to AoS. 
Reference \cite{Sun:2018:SPAWC} proposes to use the mutual information between the real-time source value and the delivered samples at the receiver to quantify the freshness of the information contained in the delivered samples. It shows that for a time-homogeneous Markov chain, the mutual information can be expressed as a non-negative and non-increasing function of the age.
In \cite{singh2019optimal}, a metric called ``Value of Information" (VoI) is introduced to facilitate packet scheduling for low-error Kalman filter based estimation. VoI depends on the age of the packet, as well as the mutual information between the packet content and the system status, which is equivalent to the variance of the noise associated with the measurement. 
Generally speaking, how to define a universal information freshness metric that accounts for dynamically evolving system states remains open.

In this paper, we propose an information theoretic measure of information freshness by taking the dynamics of the monitored system into account, where we introduce a two-dimensional discrete-time Markov chain to model the underlying state changes.
Our main contributions are three-fold: 

First, the introduced information theoretic measure generalizes the definition of AoI. It takes the age of updates and the dynamics of the monitored system into the definition, and suggests a unified approach to define proper age penalty functions~\cite{Kosta:2017:nonlinear} for various dynamic systems. 

Second, we establish fundamental relationship between AoI and the expected detection delay of status changes under any state-independent online updating policies. Such relationship validates the critical role of AoI in real-time status monitoring when the system evolution is stationary. It enables us to safely decouple the properties of the underlying system from the design of age-optimal sampling, scheduling and coding policies, without compromising the effectiveness of the delivered updates on  tracking status changes of the system. 

Third, we show that the generalized AoI is an affine function of the expected detection delay in an adapted Bayesian change point detection setting. This observation suggests that GAoI is a proper measure of information freshness even when the underlying system evolution is not stationary. 



\section{System Model}
\begin{figure}
	\centering
\includegraphics[width=3.5in]{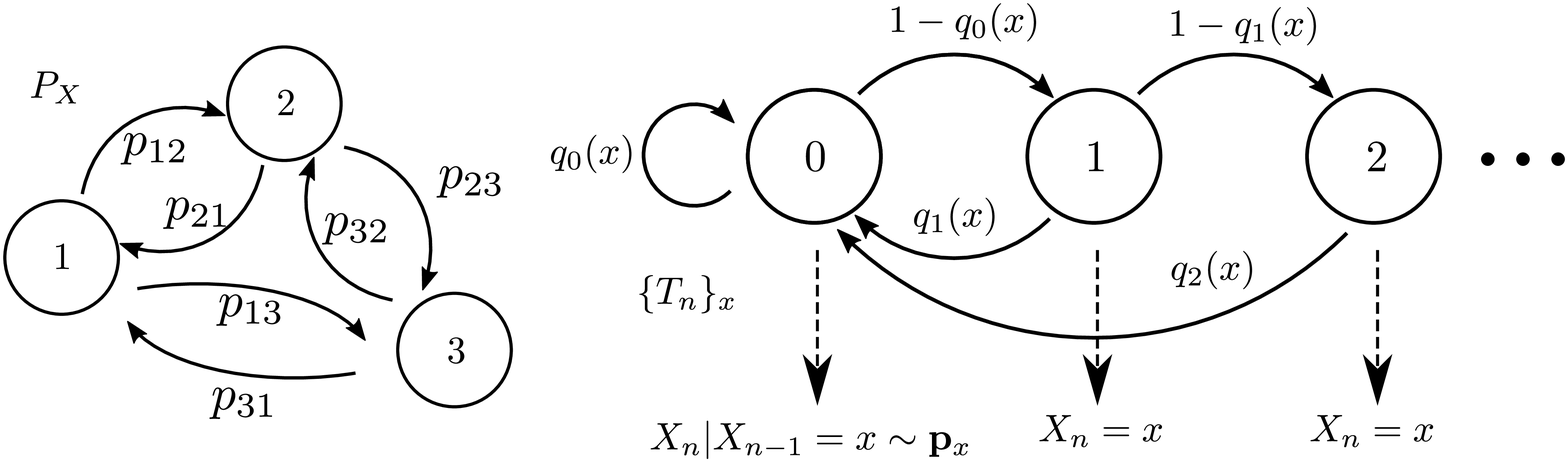}
	\caption{An {example of} the discrete-time state transition model where $\Xc=\{1,2,3\}$.}
	\label{fig:markov}
\end{figure}

Consider a single-node status monitoring scenario where a sensor monitors the status of an underlying system and sends updates to a destination through a communication channel.
In order to capture the dynamics of the monitored system, we consider a discrete-time model, where the status of the system in each time slot $n$ is denoted as $X_n$. We assume $X_n$ takes values from a {\it finite} alphabet $\Xc$. At the beginning of time slot $n$, $X_n$ may stay the same as $X_{n-1}$, or change to a different value. 
\if{0}
 \begin{figure}[t]
\centering
\includegraphics[width=3.3in]{graph/model.eps}
\caption{{System model.} }
\label{fig:model}
\end{figure} 
\fi
 We let $T_n\in \{0,1,2,\ldots\}$ denote the number of time slots the system has stayed in the current status. 
 Then, 
 \begin{align*}
 T_n=\left\{\begin{array}{ll}
 T_{n-1}+1, & \mbox{ if } X_n= X_{n-1},\\
 0, & \mbox{ if } X_n\neq X_{n-1}.\end{array}\right.
 \end{align*}
We assume the evolution of $\{T_n\}$ depends on the current status $X_n$ in general. Specifically, we denote $q_i(x):=\Pb[T_{n+1}=0|T_n=i,X_n=x]$, and denote the corresponding transition matrix as $P_T(x)$. 
Besides, if $T_n=0$, i.e., a status change happens at the beginning of time slot $n$, $X_n$ evolves according to a Markov chain with transition matrix $P_X$, as illustrated in Fig.~\ref{fig:markov}. Denote the status of the system at time $n$ as $U_n:=(X_n,T_n)$. We can show that
$U_n$ evolves according to a two-dimensional Markov chain. Essentially, such a model accommodates the scenario where $\{X_n\}$ is Markovian, as well as certain scenarios where $\{X_n\}$ cannot be simply modeled as a Markov chain, thus is more general in modeling dynamic status changes. 

\if{0}
\begin{align}
&\Pb[(X_{n+1},T_{n+1})=(x,t)|\{(X_k,T_k)\}_{k=1}^n=\{(x_k,t_k)\}_{k=1}^n] \nonumber \\
&=\Pb[(X_{n+1},T_{n+1})=(x,t)|(X_n,T_n)=(x_n,t_n)],
\end{align}
\fi

Assume each state update sampled at time $n$ is a time-stamped tuple $U_n:=(X_n,T_n)$, i.e., it does not just contain the time-stamped status $X_n$, but also the duration that the system has stayed in the current status. 
The receiver tracks the system status based on received updates.
The objective of this work is to investigate the fundamental impact of information freshness on the timely detection of status changes. As a first step, we focus on state-independent online policies defined as follows.

\begin{Definition}[State-independent online updating policies]
Let $s_1,s_2,\ldots\in\Nb$ be the sampling time points and $d_1,d_2,\ldots\in\Nb$ be the corresponding delivery times of the updates $U_{s_1}$, $U_{s_2}$, $\ldots$ under an updating policy, where we let $d_i=\infty$ if the $i$th update is never delivered. If $s_i$ only depends on previous sampling points $\{s_j\}_{j=1}^{i-1}$ and up-to-date delivery times {$\{d_j: d_j\leq s_i\}$}, and $d_i$s are independent with $\{U_n\}$, the policy is called a state-independent online updating policy. 
\end{Definition}

\begin{figure}[t]
\centering
\includegraphics[width=3.2in]{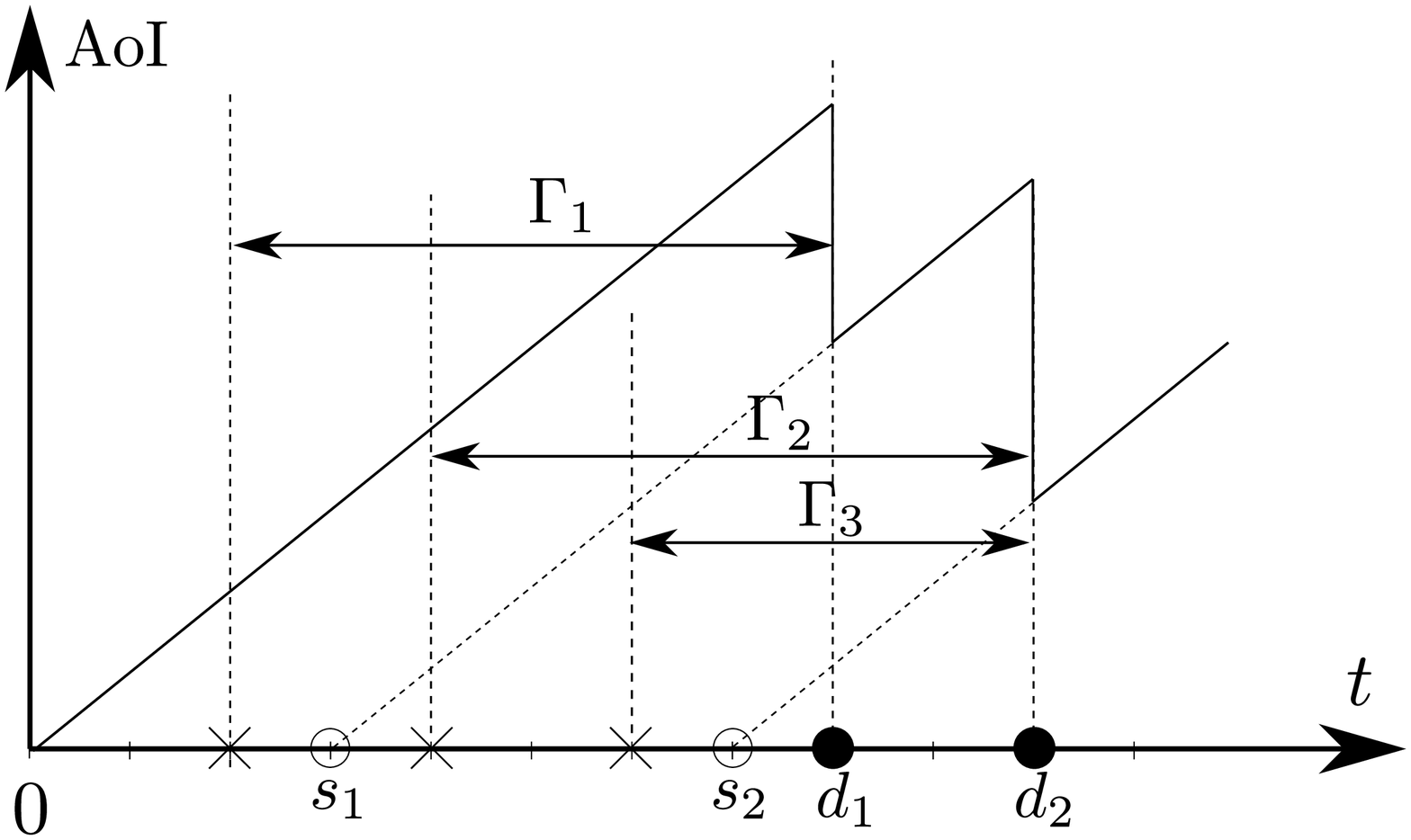}
\caption{\small{AoI evolution with given status change points. $\times$ represents status changes, $\Circle$ represents sampling points, and solid circle represents delivery time of an update.  $\Gamma_i$ is the detection delay for the $i$th status change.}}
\label{fig:delay}
\end{figure}

We also formally define detection delay as follows.
\begin{Definition}[Detection delay of status changes]
If the system status changes at the beginning of time slot $n$, i.e., $T_n=0$, the detection delay of this status change equals
$\min \{d_i: s_i \geq T_n\}-T_n$.
\end{Definition}
The definition of detection delay is intuitive in the sense that a status change won't be detected until a status update collected after the change is delivered to the destination for the first time. From an information theoretic perspective, if multiple status changes happen between the sampling points of two consecutively delivered updates, as shown in Fig.~\ref{fig:delay}, the destination won't be able to {\it locate} all of them except the last one based on $U_{\delta(n)}$ only. However, we still consider them to be {\it detected} in order to properly measure the detection delay performance of the updating schemes.

\section{Information Theoretic Characterization of Information Freshness}

Assume the latest status update at the destination at time $n$ was sampled at time $\delta(n)$. Then, the instantaneous AoI at the destination is $n-\delta(n)$. 
We propose the following metric as an information staleness measure at the destination at time $n$: 
\begin{align}\label{eqn:GAoI}
\Phi(n):=H(U^{\delta(n)+1:n}|U_{\delta(n)}),
\end{align}
where $U^{n:n+k}:=\{U_{n},U_{n+1},\ldots,U_{n+k}\}$.
Intuitively, $\Phi(n)$ captures the ``novelty" in random process $\{U_n\}$ since time $\delta(n)$, which also indicates the uncertainty level at the destination regarding the monitored system status over duration $[\delta(n)+1,n]$. We term it as {\it generalized Age of Information} (GAoI) in this paper.

The GAoI defined in (\ref{eqn:GAoI}) {essentially} measures information freshness from an {ensemble's} perspective, i.e., it averages over all possible realizations of $\{U_n\}$. Therefore it is {\it state-independent}, i.e., it is independent with the information content of the received updates (i.e., $U_{\delta(n)}$); Rather, it only relies on stochastic information of the state evolution process (i.e., $\{P_T(x)\}$ and $P_{X}$), as well as the time stamps of received updates. 

Intuitively, the information contained in $U_{\delta(n)}$ may indicate how fast the system status may change from the previous status, and how much uncertainty/novelty is generated since the latest update was generated. This motivates the definition of {\it state-dependent} GAoI as follows. 

Define 
\begin{align}\label{eqn:GAoI2}
  \Phi(n|U_{\delta(n)}=u)=H(U^{\delta(n)+1:n}|U_{\delta(n)}={u}),
\end{align}
i.e., the uncertainty in system states over $[\delta(n)+1,n]$ given the latest observation $U_{\delta(n)}=u$. 

Then, based on the definition of conditional entropy, we have
\begin{align}
    \Phi(n)=  \sum_u\Phi(n|U_{\delta(n)}=u)\cdot\Pb[U_{\delta(n)}=u]. 
\end{align}

The discrete-time status evolution model enables us to explicitly discuss the relationship between GAoI, AoI and the detection delay of status changes. In the following, we investigate two scenarios, and establish their relationships.

\section{Stationary Markovian State Evolution }\label{sec:state_indep}
In this section, we focus on stationary Markovian state evolution models. 


\begin{Proposition}
\label{prop:AoI}
Assume $(X_n,T_n)$ evolves according to a {stationary Markov chain} $\Cc$ defined by transition matrices $P_X$ and $\{P_T(x)\}_{x\in\Xc}$.
Then,
\begin{align*}
\Phi(n)
&=(n-\delta(n))\cdot H(P_X, \{P_T(x)\}_{x\in\Xc}),
\end{align*}
where $H(P_X, \{P_T(x)\}_{x\in\Xc})$ is the entropy rate of $\Cc$.
\end{Proposition}
Due to space limitation, we omit the proofs of propositions and corollaries in this paper. 
Proposition~\ref{prop:AoI} can be proved based on the property of stationary Markov chains. 
It indicates that the information freshness measure $\Phi(n)$ is proportional to the instantaneous age $n-\delta(n)$ if $H(P_X, \{P_T(x)\}_{x\in\Xc})\neq 0$, which validates the effectiveness of AoI in capturing uncertainty in the system status from the destination's perspective. 

Besides, compared with AoI which only depends on time, $\Phi(n)$ also captures the dynamics of the underlying status evolution. 
For systems with frequent state changes, or with very dynamic state transitions, $\Phi(n)$ grows quickly as age increases, which implies that the information gets stale quickly. On the other hand, in systems with more deterministic state changes, information will ``age" at a slower rate. 

One extreme case is when the status of the system evolves in a periodic pattern, i.e., given an initial status $X_0$, the system evolves according to a cycle over the states in $\Xc$. Thus, $\Phi(n)=0$, i.e., $U_{\delta(n)}$ will never get expired, since it can be used to accurately predict the state of the system in any upcoming time. Therefore, AoI itself cannot be used for a measure of information freshness for this scenario.  This special case illustrates the importance of taking the dynamics of status evolution into the definition of information freshness.

\if{0}
\begin{Proof}
By the definition $\Phi(n)$, we have
\begin{align}
&\Phi(n)=H(U^{\delta(n)+1:n}|U_{\delta(n)}) \nonumber \\
&=\sum_{i=\delta(n)+1}^{n}H(U_i|U^{\delta(n):i-1})=\sum_{i=\delta(n)+1}^{n}H(U_i|U_{i-1}) \label{eqn:prop1:1} \\
&=(n-\delta(n))\cdot H(P_X, \{P_T(x)\}_{x\in\Xc})\label{eqn:prop1:3}
\end{align}
where Eqn.~(\ref{eqn:prop1:1}) follows from the assumption that $\{U_n\}$ evolves according to a stationary Markove chain. 
\end{Proof}
\fi


\begin{Proposition}
Assume the stationary distribution of $(X_n,T_n)$ exists, and denote it as $\{\mu_{x,i}\}_{x\in\mathcal{X},i\in \Zb_+}$, where $\mu_{x,i}=\Pb[(X_n,T_n)=(x,i)]$. 
Then, $H(P_X,\{P_T(x)\}_{x\in\mathcal{X}})$ equals 
\begin{align*}
\sum_{x\in\mathcal{X}}\hspace{-0.03in}\mu_{x,0}\hspace{-0.03in}\sum_{i=0}^\infty \hspace{-0.03in}\prod_{j=0}^{i-1}(1\hspace{-0.03in}-\hspace{-0.03in}q_j(x))[H(q_{i}(x),1\hspace{-0.03in}-\hspace{-0.03in}q_{i}(x))\hspace{-0.03in}+\hspace{-0.03in}q_{i}(x)H(\mathbf{p}_x)],
\end{align*}
where $\mathbf{p}_x$ is the row associated with status $x$ in $P_X$, and $H(\pi)$ is the entropy of distribution $\pi$.
\end{Proposition}

\begin{Corollary}
If $P_T(x)$ is homogeneous, i.e., $P_T(x)=P_T$ for all state $x\in\mathcal{X}$, then, the entropy rate of the Markov chain $\Cc$ equals
\begin{align}
H(P_X,P_T)=H(P_T)+H(P_X)\Pb(T_n=0).
\end{align}
\end{Corollary}

Our main result is summarized as follows. 



\begin{Theorem}
\label{thm:stationary}
Assume the monitor is informed of the initial system state at time 0. Then, under any state-independent online updating policy,  $$\frac{\bar{\Phi}(T)}{H(P_X,\{P_T(x)\})} =\bar{\Delta}(T)=\frac{\bar{\Gamma}(T)}{\Pb(T_n=0)},$$ where $\bar{\Phi}(T):=\sum_{n=1}^T\Phi(n)$, $\bar{\Gamma}(T)$ is the expected total detection delay of the state changes over $[1,T]$, and $\bar{\Delta}(T)$ is the expected total AoI experienced at the destination over $[1,T]$.
\end{Theorem}

\if{0}
\begin{Proposition}[Relationship with Detection Delay]\label{prop:detection_delay}
Under any updating (i.e., sampling and transmission) schemes,  $$\sum_{n=1}^T\Phi(n)=H(P_X) \bar{\Gamma} +H(P_T)\bar{\Delta},\quad \bar{\Gamma} =\bar{\Delta}\cdot \Pb(T_n=0),$$ where $\bar{\Gamma}$ is the expected total detection delay of the state changes over $[1,T]$, and $\bar{\Delta}$ is the expected total AoI experienced at the destination over $[1,T]$.
\end{Proposition}
\fi
\begin{Proof}
\if{0}
By Proposition~\ref{prop:AoI}, we have the following result
\begin{align}
\sum_{n=1}^{T}\Phi(n)&=H(P_X)\bar{\Delta}\cdot\Pb(T_n=0)+H(P_T)\bar{\Delta}, \label{eqn:delay:1}
\end{align}
where $\bar{\Delta}$ is the expected total AoI experienced at the destination over $[1,T]$. We are left to show the expected total detection delay of the state changes over $[1,T]$ is exactly $\bar{\Delta}\cdot\Pb(T_n=0)$. 
\fi
Consider the sampling times and update delivery times under a state-independent online updating policy.  
For ease of exposition, let $s_i$ and $d_i$ be the $i$th sampling time and the corresponding delivery time. Consider the first $T$ time slots $[1,T]$ during which $K$ updates are delivered. We collectively denote the delivery times as $\{d_i\}_{i=1}^K$ and the corresponding sampling times as $\{s_i\}_{i=1}^{K}$. We also define {$d_0=s_0=0$} and $d_{K+1}=s_{K+1}=T$. Without loss of generality, we can assume $d_i< d_j$ whenever $s_i<s_j$. In fact, if there exists $s_j\geq s_i$ such that $d_j\leq d_i$, the information delivered at time $d_i$ is stale compared to the previously received information $U_{s_j}$. Thus, we can exclude such updates without affecting the (G)AoI evolution or detection delay. Define the detection delay of a status change at time $n$ $(n\leq T)$ restricted to $[1,T]$ as $\Gamma=\min \{d_i: s_i\geq n, i\leq K+1\}-n$. The expected total detection delay over $[1,T]$ can be expressed as
\begin{align}
\Bar{\Gamma}(T)&=\Pb(T_n=0)\sum_{i=0}^{K}\sum_{j=s_i+1}^{s_{i+1}}(d_{i+1}-j),
\end{align}
where the equality is based on the fact that $\Pb(T_n=0)$ is a constant for all $n$ under the assumption that $\{U_n\}$ is a stationary Markov chain. Then, the expected total detection delay over $[1,T]$ scaled by $1/\Pb(T_n=0)$ is
\begin{align}
\frac{\bar{\Gamma}(T)}{\Pb(T_n=0)}
&=\frac{1}{2}T^2-\frac{1}{2}T-\sum_{i=1}^{K}s_i(d_{i+1}-d_{i}).
\end{align}
The expected AoI experienced over $[1,T]$ is
\begin{align}
\bar{\Delta}(T)&={\sum_{j=0}^{K+1}\sum_{i=0}^{d_{j+1}-d_j-1}(d_j-s_j+i)} \\
&=\frac{1}{2}T^2-\frac{1}{2}T-\sum_{i=1}^{K}s_i(d_{i+1}-d_i).
\end{align}
Hence, $\bar{\Gamma}(T)=\bar{\Delta}(T)\cdot\Pb(T_n=0)$. We note that this equation holds for any possible realizations of $s_i$s and $d_i$s under a state-independent online updating policy. Thus, it still holds if we take expectations with respect to $s_i$s and $d_i$s. The first equality then follows from Proposition~\ref{prop:AoI}.
\end{Proof}


Theorem~\ref{thm:stationary} validates the fundamental role of information freshness on timely detection of status changes. It indicates that minimizing AoI through state-independent sampling, transmission scheduling or coding is equivalent to minimizing the expected detection delay of the status changes in the system. 


\if{0}
\begin{Corollary}[Relationship with Network Latency]\label{prop:delay}
Consider an updating protocol where the source will generate an update to the destination whenever the state changes (i.e., when $T_n=0$), and all updates are delivered in an First-In-First-Out fashion. Then, $$\sum_{n=1}^T\Phi(n)=H(P_X) \bar{D} +H(P_T)\bar{\Delta},\quad \bar{D} =\bar{\Delta}\cdot \Pb(T_n=0),$$ where $\bar{D}$ is the expected total delay experienced by the update packets over $[0,T]$.
\end{Corollary}

Corollary~\ref{prop:delay} is an importance observation magnifying the intricate relationship between information freshness and network delay. It indicates that under {\it event-triggered} status update generation scheme and FIFO transmission protocol, network latency can be directly translated to information staleness at the destination. However, if the sampling and update generation are not event-triggered, or the transmission are not FIFO, network latency is in general not a proper measure of information freshness.
\fi

\section{Non-stationary Markovian State Evolution}
The result in Section~\ref{sec:state_indep} relies on the stationary Markovian state evolution assumption.  It is also intuitive that under the stationary Markovian assumption, $\Phi(n)$ scale proportionally to the instantaneous age at the destination. However, such assumptions are quite restrictive in practice. This motivates us to investigate a broader class of status change models where the stationary assumption may not hold. While this seems challenging for a general setting, as a first step, we consider the following status change model, which is adapted from the classical Bayesian change point detection model in the literature.

We assume the system status starts with an initial state $0$ at $n=0$. At an unknown change point $\theta\in\Nb$, the system status changes to $1$. Under the Bayesian setting, it assumes that $\theta$ is a random variable following a geometric distribution, i.e., $\Pb[\theta=\tau]=p(1-p)^{\tau-1}$, $\tau=1,2,\ldots$.

\begin{figure}[t]
	\centering
\includegraphics[width=2.5in]{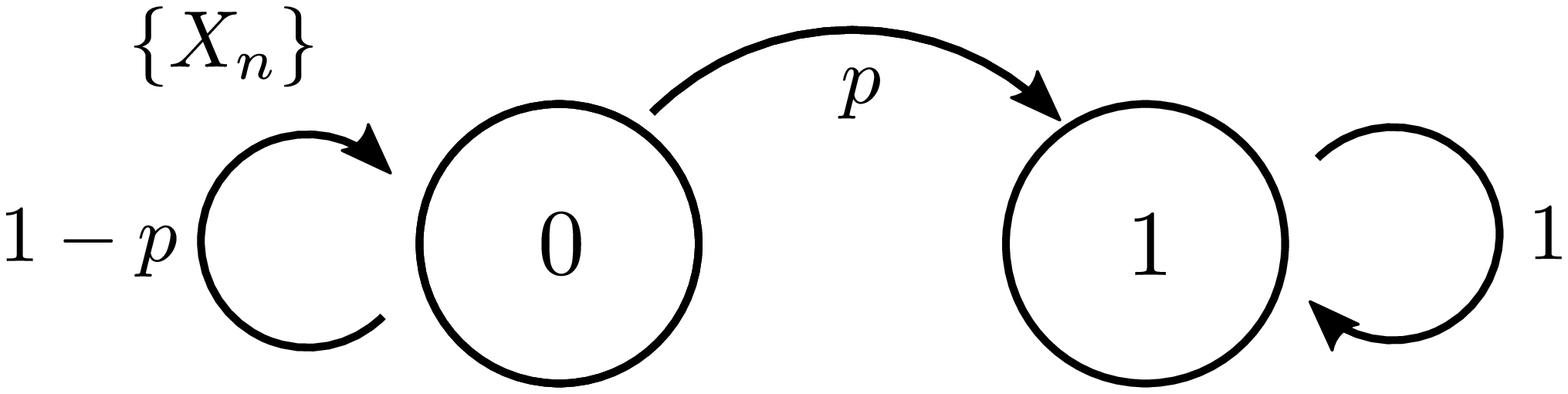}

	\caption{Equivalent Markov chain for the Bayesian change point model.}
	\label{fig:changepoint_model}
\end{figure}

We can show that the system status $\{X_n\}$ can be modeled as a Markov chain shown in Fig.~\ref{fig:changepoint_model}. Since it has an absorbing state 1, and the initial state $X_0=0$, the Markov chain is {no longer stationary}. 

Since the state evolution can be characterized by {a one-dimensional} Markov chain $\{X_n\}$, the information theoretic definitions of information freshness in Eqns. (\ref{eqn:GAoI}) and (\ref{eqn:GAoI2}) are equivalent to the following definitions, respectively:
\begin{align}
& \Phi(n)=H(X^{\delta(n)+1:n}|X_{\delta(n)}),\\
 &   \Phi(n|X_{\delta(n)}=x)=H(X^{\delta(n)+1:n}|X_{\delta(n)}=x).
\end{align}

We have the following observations.

\begin{Proposition}\label{prop:bayesion}
Under the Bayesian change point model, 
\begin{align*}
    \Phi(n|X_{\delta(n)}=0)&=0,\\
    \Phi(n|X_{\delta(n)}=1)&=H(p,\hspace{-0.02in}\cdots\hspace{-0.02in},(1\hspace{-0.02in}-\hspace{-0.02in}p)^{n-\delta(n)-1}p, (1\hspace{-0.02in}-\hspace{-0.02in}p)^{n-\delta(n)}),
\end{align*}
where $H(p_1,p_2,\cdots,p_n)=-\sum_{i}p_i\log{p_i}$. 
\end{Proposition}



\if{0}
\begin{align}
&\Phi(n|X_{\delta(n)}=x)=H(U^{\delta(n)+1:n}|X_{\delta(n)}=x) \nonumber \\
&=H(T^{\delta(n)+1:n}|X_{\delta(n)}=x)+H(X^{\delta(n)+1:n}|T^{\delta(n)+1:n},X_{\delta(n)}=x)\nonumber \\
&=H(T^{\delta(n)+1:n}|X_{\delta(n)}=x), \label{eqn:ex2:1}
\end{align}
where the last equality follows from the fact that $X^{\delta(n)+1:n}$ is uniquely determined by $T^{\delta(n)+1:n}$ and $X_{\delta(n)}=x$.

Clearly, $\Phi(n|X_{\delta(n)}=1)=0$. We will focus on the case where $X_{\delta(n)}=0$. We have
\begin{align}
&\Phi(n|X_{\delta(n)}=0)=H(T^{\delta(n)+1:n}|X_{\delta(n)}=0) \\
&=H(p;(1-p)p;\cdots;(1-p)^{n-\delta(n)-1}p;(1-p)^{n-\delta(n)}), \label{eqn:ex2:2}
\end{align}
where $H(p_1;p_2;\cdots;p_n)=\sum_{i}p_i\log\frac{1}{p_i}$ is the entropy of probability distribution $\{p_1;p_2;\cdots;p_n\}$ where $\sum_i p_i=1$. 
\fi

\begin{figure}
	\centering
\includegraphics[width=3in]{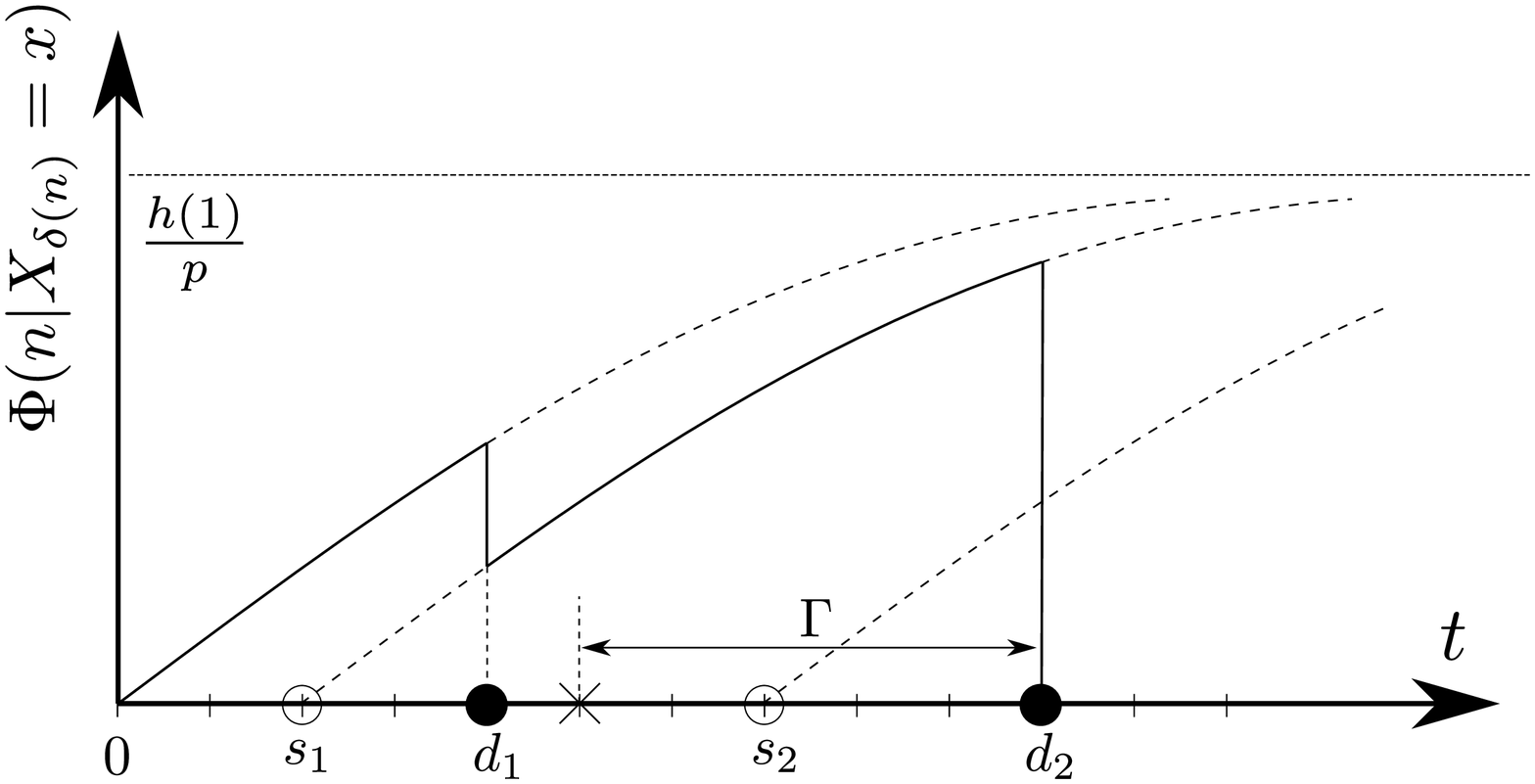}

	\caption{\small{Generalized AoI evolution with given sampling times $s_i$s and delivery times $d_i$. $\times$ represents the change point, and $\Gamma$ represents the detection delay.}}
	\label{fig:changepoint_sample}
	\vspace{-0.1in}
\end{figure}

\begin{Theorem}\label{thm:change_detection}
Under any state-independent online updating policy,  $$\bar{\Phi}(T)=\frac{H(p,1-p)}{p}\bar{\Gamma}(T)+C(T),$$ where $\bar{\Phi}(T):=\sum_{n=1}^T\Phi(n)$, $\bar{\Gamma}(T)$ is the expected detection delay of the state change over $[1,T]$, and $C(T)$ is a constant for fixed $T$. 

\end{Theorem}
\begin{Proof}
Denote $h(x)$ as
\begin{align}
h(x):=H(p,\cdots,(1-p)^{x-1}p,(1-p)^x), \forall x\in\Nb.
\end{align}
Based on the definition of entropy, we have the following recursive formula:
\begin{align}
h(x+1)=h(x)+(1-p)^x h(p).
\end{align}
Denote $h(1):=H(p,1-p)$. Then,
\begin{align}
h(x)=\frac{1-(1-p)^x}{p}h(1). \label{eqn:ex2:3}
\end{align}
Plugging (\ref{eqn:ex2:3}) into the expression in Proposition~\ref{prop:bayesion}, we have
\begin{align}
\Phi(n|X_{\delta(n)}=0)=\frac{1-(1-p)^{n-\delta(n)}}{p}h(1).
\end{align}
Then, a uniform expression for the state-dependent GAoI is as follows: 
\begin{align}
& \Phi(n|X_{\delta(n)}=x)\nonumber\\
& =\frac{1-(1-p)^{n-\delta(n)}}{p}h(1)\lv_{X_{\delta(n)}=0}, \quad x\in\{0,1\}.
\end{align}


Similar to the proof of Theorem~\ref{thm:stationary}, we consider the sampling points $s_1,s_2,\ldots,s_K$ and corresponding delivery points $d_1,d_2,\ldots,d_K$ of the $K$ updates delivered over $[1,T]$ under a given state-independent updating policy, as shown in Fig.~\ref{fig:changepoint_sample}. Without loss of generality, we assume $0<s_1<s_2<\ldots<s_K<T$ and $0<d_1<d_2<\ldots<d_K<T$.
We define $s_0=d_0=0$ and $s_{K+1}=d_{K+1}=T$ for ease of exposition. 

The accumulative generalized GAoI can be expressed as
\begin{align}
\sum_{n=1}^T\Phi(n|X_{\delta(n)}=x)
\hspace{-0.02in}=\hspace{-0.02in}\sum_{i=0}^K\sum_{j=d_i+1}^{d_{i+1}}\hspace{-0.02in}\frac{1-(1-p)^{j-s_i}}{p}h(1)\lv_{X_{s_i}=0}. \nonumber
\end{align}
Taking expectation with respect to $x$, we have
\begin{align}
& \bar{\Phi}(T) \nonumber \\
&=\sum_{i=0}^K\sum_{j=d_i+1}^{d_{i+1}}\frac{1\hspace{-0.02in}-\hspace{-0.02in}(1-p)^{j-s_i}}{p}h(1)(1-p)^{s_i} \nonumber\\
&=\frac{h(1)}{p}\left(\hspace{-0.02in}\frac{(1-p)[(1-p)^T-1]}{p}
+\sum_{i=0}^{K}(d_{i+1}-d_i)(1-p)^{s_i}\hspace{-0.02in}\right).\label{eqn:phi_bayesian}
\end{align}

On the other hand, the expected detection delay can be calculated as follows: 
\begin{align}
&\bar{\Gamma}(T)\nonumber\\
&=\sum_{i=0}^{K}\sum_{j=s_i+1}^{s_{i+1}}(1-p)^{j-1}p(d_{i+1}-j) \\
&=-\sum_{k=1}^{T}k(1-p)^{k-1}p+\sum_{i=0}^{K}\sum_{j=s_i+1}^{s_{i+1}}(1-p)^{j-1}pd_{i+1}  \\
&=-\sum_{k=1}^{T}k(1-p)^{k-1}p+\sum_{i=0}^{K}d_{i+1}[(1-p)^{s_i}-(1-p)^{s_{i+1}}]  \nonumber \\
&=\hspace{-0.02in}-T(1\hspace{-0.02in}-\hspace{-0.02in}p)^T\hspace{-0.02in}-\hspace{-0.02in}\sum_{k=1}^{T}k(1\hspace{-0.02in}-\hspace{-0.02in}p)^{k-1}p+\sum_{i=0}^{K}(d_{i+1}\hspace{-0.02in}-\hspace{-0.02in}d_i)(1\hspace{-0.02in}-\hspace{-0.02in}p)^{s_i}.\label{eqn:delay_bayesion}
\end{align}

Since the updating policy is {state-independent}, taking expectation of (\ref{eqn:phi_bayesian}) and (\ref{eqn:delay_bayesion}) with respect to $s_i$s, we have the proof complete.
\end{Proof}

\if{0}
\begin{align}
&\sum_{n=1}^{T}\Phi(n|X_{\delta(n)}=x) \nonumber \\
&=\sum_{n=1}^{m}\frac{1-(1-p)^n}{p}h(1)+\sum_{n=m+1}^{T}\frac{1-(1-p)^{n-m}}{p}h(1)\lv_{X_m=0} \nonumber \\
&=\frac{h(1)}{p}\left(m-\frac{1-p}{p}[1-(1-p)^{m}]\right) \nonumber \\
& \quad +\lv_{X_m=0}\frac{h(1)}{p}\left(T-m-\frac{1-p}{p}[1-(1-p)^{T-m}]\right).
\end{align}
Take expectation with respect to $X^{1:T}$ and we have
\begin{align}
&\Eb[\sum_{n=1}^{T}\Phi(n|X_{\delta(n)}=x)]=\frac{h(1)(1-p)}{p^2}[(1-p)^{T}-1] \nonumber \\
&\quad +\frac{h(1)}{p}[m+(1-p)^m(T-m)],\quad \mbox{if $m\in[1:T]$}
\end{align}
If $m=\varnothing$, the expected summed generalized AoI is
\begin{align}
&\Eb[\sum_{n=1}^{T}\Phi(n|X_{\delta(n)}=x)]=\sum_{n=1}^{T}\frac{1-(1-p)^n}{p}h(1) \\
&=T-\frac{1-p}{p}[1-(1-p)^{T}].
\end{align}

The expected detection delay during $[1,T]$ is straightforward \tcr{detection occurs at $t=m+0^+$}
\begin{align}
&\bar{\Gamma}=\sum_{n=1}^{T}(1-p)^{n-1}p[(m-n)\lv_{n\leq m}+(T-n)\lv_{n> m}] \\
&=\sum_{n=1}^{T}(1-p)^{n-1}p(T-n)+\sum_{n=1}^{m}(1-p)^{n-1}p(m-T) \nonumber\\
&=\sum_{n=1}^{T}(1-p)^{n-1}p(T-n)+(m-T)[1-(1-p)^{m}] \nonumber  \\
&=C(T)-T+m+(1-p)^{m}(T-m), \quad  \mbox{if $m\in[1:T]$}
\end{align}
where $C(T)=\sum_{n=1}^{T}(1-p)^{n-1}p(T-n)$ is a constant if $T$ is fixed and
\begin{align}
&\bar{\Gamma}=\sum_{n=1}^{T}(1-p)^{n-1}p (T-n)=C(T),\quad \mbox{if $m=\varnothing$}.
\end{align}

In this toy example, we showed that the relation between state-dependent generalized AoI and the total detection delay. Observe that the AoI minimization is to choose the midpoint, which is different from minimizing the generalized AoI and the total detection delay.
\fi

\if{0}
First, we will consider the case when each row of $P_X$ is a permutation of other rows. Thus, $H(X_n|X_{n-1}=x,T_n=0)$ becomes a constant for any $x\in\Xc$. Then, the state-dependent generalized AoI only depends on $\delta(n)$,  $T_{\delta(n)}$, as well as $P_X$ and $P_T$. We expect that by isolating the impact of $X_{\delta(n)}$ from the formulation, closed-form expressions of state-dependent generalized AoI are attainable. We will also study its relationship with state change detection delay for this case. Next, we will isolate the impact of $T_{\delta(n)}$ from the picture, and focus on the case where $T_n=0$ for every $n$, i.e., the state is constantly changing. We expect to obtain closed-form expressions for certain classes of $P_X$. With insights obtained from those special cases, we will then investigate more general models.

\subsection{tbc}
\begin{Proposition}
\tcr{incorrect, check proposition 1}
\begin{align}
\Phi(n|U_{\delta(n)}=[x,0])=
\end{align}
\end{Proposition}
\fi

Theorem~\ref{thm:change_detection} indicates that GAoI is an affine function of the expected detection delay for the Bayesian change point model under any state-independent updating policy. We can verify that such relationship no longer holds between the AoI and the expected detection delay. Such result suggests that GAoI is a more accommodating measure of information freshness compared with AoI. 

\section{Simulation Results}

\begin{figure}
	\centering
\includegraphics[width=3in]{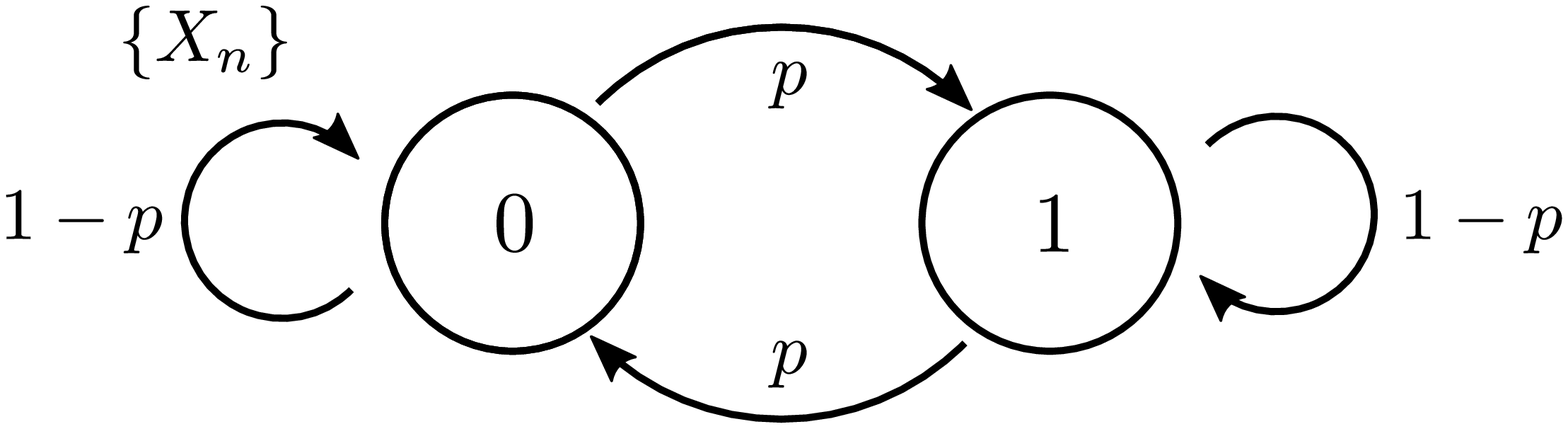}
	\caption{Generalized AoI evolution with given sampling points.}
	\label{fig:simu1_model}
	\vspace{-0.1in}
\end{figure} 

In this section, we evaluate our results through simulations. We evaluate the AoI, GAoI and the status change detection delay under two state evolution models: a stationary two-state symmetric Markov chain and a Bayesian change point model.

\subsection{Stationary Two-state Symmetric Markov chain}\label{sec:simu1}
\begin{figure}
\vspace{-0.1in}
\centering
\includegraphics[width=3.3in]{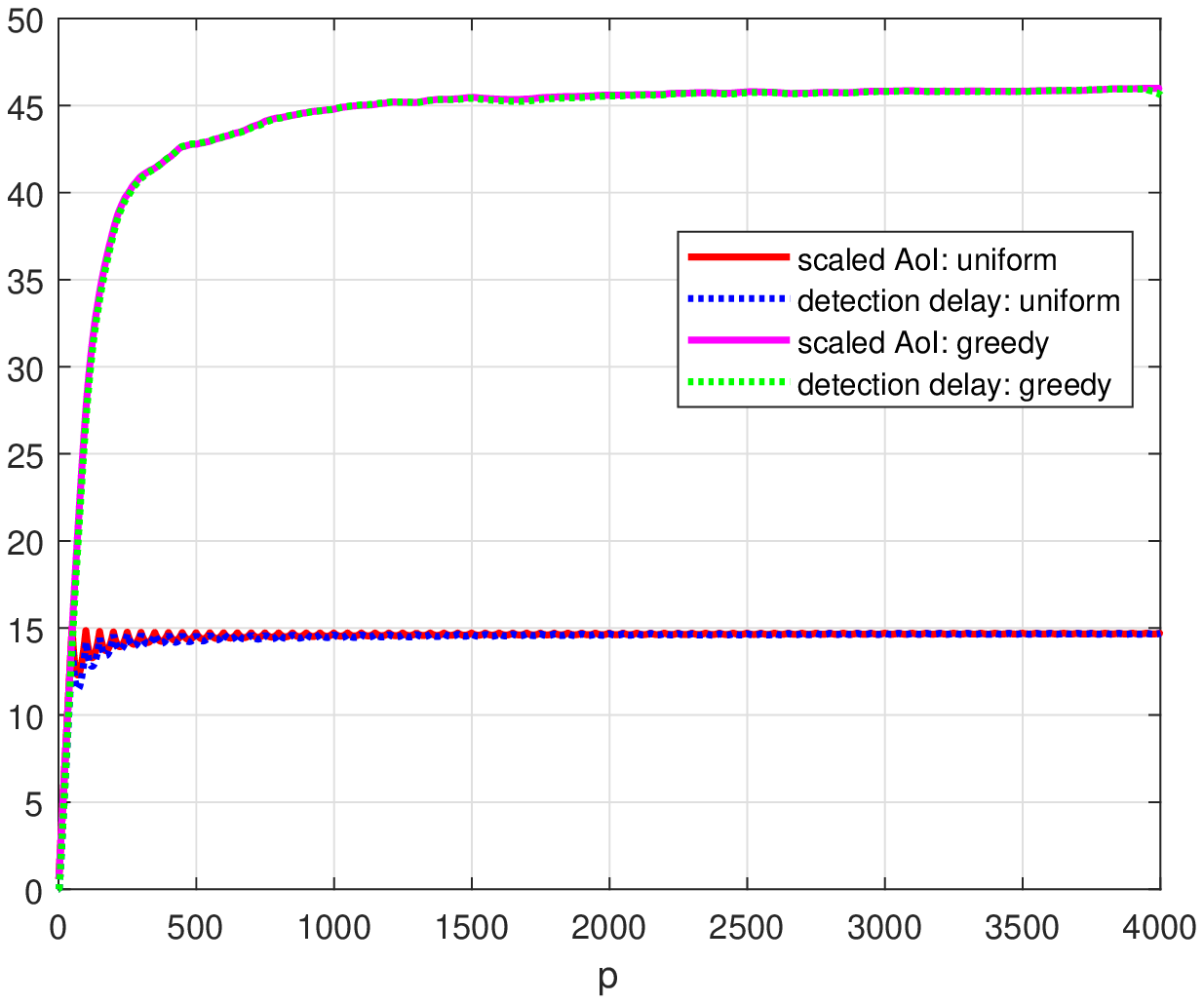}
\caption{AoI and detection delay for a stationary Makovian status evolution model.}
\vspace{-0.1in}
\label{fig:simu1}
\end{figure}

First, we consider a two-state symmetric Markov chain where the probability of status change is $p$, as illustrated in Fig.~\ref{fig:simu1_model}. We adopt two different updating schemes: The first one is uniform sampling with instant delivery where the sampling occurs at $n=50,100,\ldots$. The other one is greedy sampling policy with random delivery time, where the sampling happens whenever an update is delivered to the destination. We assume the delivery time of each update is uniformly distributed in $[20,80]$. We note that the sampling rates of the two schemes are actually the same, and both policies are state-independent. In this simulation, we fix $p=0.6$. For each updating policy, we generate $100$ sample paths, where the initial state is randomly selected according to the stationary distribution. For each sample path, we track the status changes and obtain the total detection delay, which is then averaged over $T$. We also track the AoI evolution and calculate its time average. As shown in Fig.~\ref{fig:simu1}, after scaling AoI by a factor of $\Pb(T_n=0)$, the ensemble average matches the ensemble average of the detection delay under both updating policies. This is consistent with the theoretical results in Theorem~\ref{thm:stationary}.

\subsection{Bayesian Change Point Model}
\begin{figure}
\vspace{-0.1in}
\centering
\includegraphics[width=3.3in]{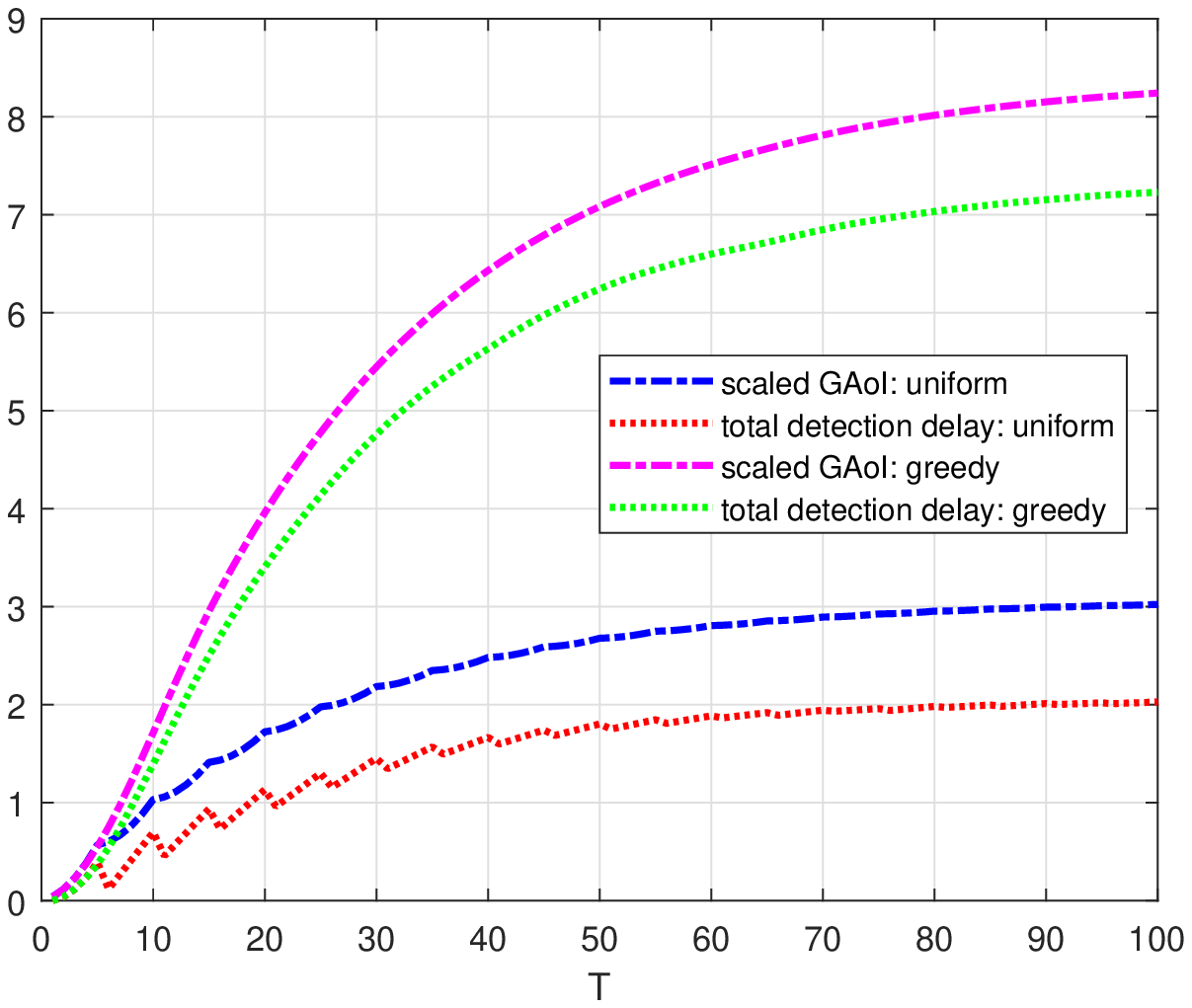}
\caption{GAoI and detection delay for a Bayesian change point model.}\vspace{-0.1in}
\vspace{-0.1in}
\label{fig:simu2}
\end{figure}

Next, we evaluate the relationship between the GAoI and the  detection delay under a Bayesian change point model. We set $p=0.04$ and track the detection delay and the evolution of the state-dependent GAoI for each sample path under the two updating policies described in Sec.~\ref{sec:simu1}. We fix the sampling rate for the uniform sampling as once every five time slots. For the other policy, we let the random delivery time be uniformly distributed in $[2,8]$. We generate $2000$ sample paths over duration $[1,100]$. The ensemble average of the cumulative GAoI and the ensemble average of the detection delay are compared and plotted in Fig.~\ref{fig:simu2}. The GAoI curves are scaled by a factor of {$\frac{p}{H(p,1-p)}$}, as suggested in Theorem~\ref{thm:change_detection}. 
It is noteworthy that the difference between the scaled GAoI and the detection delay is a constant under both updating policies, which corroborates the theoretical result in Theorem~\ref{thm:change_detection}.


\if{0}

In this section, we evaluate our results through simulations. We perform two simulations, one for , the other one for

\subsection{AoI and Detection Delay}
Consider two-state symmetric Markov chain where the probability of state change is $p$, as illustrated in~Fig.\tcr{add}. We adopt uniform sampling scheme where sampling occurs at $t=50,100,\ldots$. We examine different distribution of the delivery delay, instantaneous delivery and uniform delivery delay. For the case of uniform delivery delay, we set the delivery delay distributed uniformly over $[30:80]$\tcr{check programming}. In this scenario, it is clear that the sampling process does not behave in a first-come-first-serve (FCFS) manner. We vary $p$ within range $[0.2,0.8]$ and for each $p$, we generate $100$ samples for each pattern.

\subsection{Bayesian change point detection}

\fi
\if{0}

\subsection{State-independent Updating}
Consider a stationary Markov chain model where the state transition follows the equation:
\begin{align}
X_{n+1}
\left\{
\begin{array}{ll}
=X_n, & \mbox{with prob. } p , \\
\neq X_n , & \mbox{with prob. } 1-p,
\end{array}\right.
\end{align}
or equivalently
\begin{align}
T_{n+1}=
\left\{
\begin{array}{ll}
T_n, & \mbox{with prob. } p , \\
0 , & \mbox{with prob. } 1-p.
\end{array}\right.
\end{align}
We adopt uniform sampling strategy and assume instantaneous delivery of the sampled updates. The sampling rate is $1/N$ where $N$ is an integer. Without loss of generality, assume the sampling occurs at $t=kN$ where $k$ is an non-negative integer. \tcr{Add ref. or prove uniform sampling is optimal in term of AoI (G-AoI?) given sampling rate constraint}

Clearly, the system evolves according to a renewal process. In the following, we focus on the first renewal interval $(0,N]$ and find the average AoI $\bar{\Delta}_{avg}$ and expected average detection delay of the state changes $\bar{\Gamma}_{avg}$. The average AoI $\bar{\Delta}_{avg}$ is obvious
\begin{align}
\bar{\Delta}_{avg}=\frac{1}{N}\sum_{i=0}^{N-1}i=\frac{N-1}{2}.
\end{align}
In order to derive the expected average detection delay of the state changes $\bar{\Gamma}_{avg}$, define $\gamma_i$ where $i=1,\ldots,N$ as the detection delay of the update generated at time $t=i$ if exists and $\gamma_i=0$ otherwise. Then
\begin{align}
\gamma_i=
\left\{
\begin{array}{ll}
0, & \mbox{with prob. } p , \\
N-i , & \mbox{with prob. } 1-p,
\end{array}\right.
\end{align}
and we have
\begin{align}
\bar{\Gamma}_{avg}&=\frac{1}{N}\Eb[\sum_{i=1}^{N}\gamma_i]=\frac{1}{N}\sum_{i=1}^{N}\Eb[\gamma_i] \\
&=\frac{1}{N}\sum_{i=1}^{N}(1-p)(N-i)=\frac{(1-p)(N-1)}{2}.
\end{align}

By Eqn.~\ref{eqn:delay:1}, the average generalized AoI is
\begin{align}
\frac{1}{N}\sum_{n=1}^{N}\Phi(n)&=H(P_X)\bar{\Delta}_{avg}\Pb(T_n=0)+H(P_T)\bar{\Delta}_{avg}\\
&=H(P_X)\bar{\Delta}_{avg}(1-p)+H(P_T)\bar{\Delta}_{avg} \\
&=H(P_X)\bar{\Gamma}_{avg}+H(P_T)\bar{\Delta}_{avg},
\end{align}
where $\Pb(T_n=0)=1-p$ is based on the definition of $T_n$. The above equation is exactly what Proposition~\ref{prop:detection_delay} suggests.

\subsection{State-dependent Updating}
\fi

\section{Conclusions}
In this paper, we introduce an information theoretic measure of information freshness and investigate its relationship between AoI and detection delay of status changes in a status monitoring system. Our results validates the fundamental role of AoI in timely change detection when the underlying system states evolve according to a stationary Markov chain. It also indicates that for a special non-stationary status change model, GAoI is a more proper measure. In the future, we will investigate the relationships between those metrics under state-dependent updating policies.

\if{0}
\subsection{Example 1}
In this section, we validate the state-independent information freshness metric through an example. 

Consider a stationary Markov chain model where the state transition follows the equation:
\begin{align}
X_{n+1}
\left\{
\begin{array}{ll}
=X_n, & \mbox{with prob. } p , \\
\neq X_n , & \mbox{with prob. } 1-p,
\end{array}\right.
\end{align}
or equivalently
\begin{align}
T_{n+1}=
\left\{
\begin{array}{ll}
T_n, & \mbox{with prob. } p , \\
0 , & \mbox{with prob. } 1-p.
\end{array}\right.
\end{align}
We adopt uniform sampling strategy and assume instantaneous delivery of the sampled updates. The sampling rate is $1/N$ where $N$ is an integer. Without loss of generality, assume the sampling occurs at $t=kN$ where $k$ is an non-negative integer. \tcr{Add ref. or prove uniform sampling is optimal in term of AoI (G-AoI?) given sampling rate constraint}

Clearly, the system evolves according to a renewal process. In the following, we focus on the first renewal interval $(0,N]$ and find the average AoI $\bar{\Delta}_{avg}$ and expected average detection delay of the state changes $\bar{\Gamma}_{avg}$. The average AoI $\bar{\Delta}_{avg}$ is obvious
\begin{align}
\bar{\Delta}_{avg}=\frac{1}{N}\sum_{i=0}^{N-1}i=\frac{N-1}{2}.
\end{align}
In order to derive the expected average detection delay of the state changes $\bar{\Gamma}_{avg}$, define $\gamma_i$ where $i=1,\ldots,N$ as the detection delay of the update generated at time $t=i$ if exists and $\gamma_i=0$ otherwise. Then
\begin{align}
\gamma_i=
\left\{
\begin{array}{ll}
0, & \mbox{with prob. } p , \\
N-i , & \mbox{with prob. } 1-p,
\end{array}\right.
\end{align}
and we have
\begin{align}
\bar{\Gamma}_{avg}&=\frac{1}{N}\Eb[\sum_{i=1}^{N}\gamma_i]=\frac{1}{N}\sum_{i=1}^{N}\Eb[\gamma_i] \\
&=\frac{1}{N}\sum_{i=1}^{N}(1-p)(N-i)=\frac{(1-p)(N-1)}{2}.
\end{align}

By Eqn.~\ref{eqn:delay:1}, the average generalized AoI is
\begin{align}
\frac{1}{N}\sum_{n=1}^{N}\Phi(n)&=H(P_X)\bar{\Delta}_{avg}\Pb(T_n=0)+H(P_T)\bar{\Delta}_{avg}\\
&=H(P_X)\bar{\Delta}_{avg}(1-p)+H(P_T)\bar{\Delta}_{avg} \\
&=H(P_X)\bar{\Gamma}_{avg}+H(P_T)\bar{\Delta}_{avg},
\end{align}
where $\Pb(T_n=0)=1-p$ is based on the definition of $T_n$. The above equation is exactly what Proposition~\ref{prop:detection_delay} suggests.
\fi

\if{0}

\subsection{Remark On The State-independent Information Freshness}
With a little abuse of notation, we denote $\{X_n\}$ as the underlying Markov chain of the state changes independent of Markov chain $\{T_n\}$. The state-independent information freshness metric is well-defined for irreducible Markov chain $\{X_n\}$. We note that $\{T_n\}$ is always irreducible. In the sequel, we characterize the generalized AoI depending on whether $\{T_n\}$ and $\{X_n\}$ are aperiodic. 

The following equation holds
\begin{align}
\Phi(n)=\sum_{i=\delta(n)+1}^{n}H(T_i|T_{i-1})+H(X_i|T_i,X_{i-1}).
\end{align}

\begin{itemize}
\item aperiodic $\{T_n\}$ and $\{X_n\}$.
\end{itemize}
\begin{align}
\Phi(n)=(n-\delta(n))[H(P_T)+H(P_X)\Pb(T_i=0)].
\end{align}

\begin{itemize}
\item aperiodic $\{T_n\}$ and periodic $\{X_n\}$.
\end{itemize}
\begin{align}
\Phi(n)=(n-\delta(n))H(P_T).
\end{align}

\begin{itemize}
\item periodic $\{T_n\}$ with period $d_1$ and aperiodic $\{X_n\}$.
\end{itemize}
\begin{align}
\Phi(n)=(n-\delta(n))H(P_X)\lfloor\frac{n-\delta(n)}{d_1}\rfloor.
\end{align}

\begin{itemize}
\item periodic $\{T_n\}$ with period $d_1$ and aperiodic $\{X_n\}$ with period $d_2$.
\end{itemize}
\begin{align}
\Phi(n)=0.
\end{align}

\subsection{Remark On The State-Dependent Information Freshness}
For reducible Markov chain $\{X_n\}$, we classify the states and find equivalence classes $\{\mathcal{C}_i\}_i$. The information contained in $U_{\delta(n)}$ indicates which equivalence class the current state belongs to and hence the uncertainty of future states. We adapt the definition of $\Phi(n)$ to measure the state-dependent information freshness as follows
\begin{align}
\Phi(n|x):=H(U^{\delta(n)+1:n}|\mathcal{C}_x),
\end{align}
where $\mathcal{C}_x=\mathcal{C}_i$ if $X_{\delta(n)}\in \mathcal{C}_i$. For recurrent equivalence class $\mathcal{C}_x$, the state-dependent information freshness metric is essentially the same as the state-independent information freshness metric and we will focus on the transient equivalence class $\mathcal{C}_x$. \tcr{When calculating $\Phi(n|x)$, we always perfectly balance the distribution of $X_{\delta(n)}$ such that for all $n$, the probability distribution of $X_n|X_{n-1}$ remains the same given $X_n\in\mathcal{C}_{\delta(n)}$. I.e., when transiting to a new equivalence class, we use the balanced distribution instead of true mass-1 distribution the initial state.} For state $s\in\mathcal{C}_i$, let $p_{s}$ be the balanced probability and denote $p_{{s},\mathcal{C}_j}$ as the probability of transition from $s_i$ to another equivalence class $\mathcal{C}_j$. Define $P_i$ as
\begin{align}
P_i:=\sum_{s\in C_i}p_s\sum_{j:s\rightarrow \mathcal{C}_j, \mathcal{C}_j\not\rightarrow s} p_{s,\mathcal{C}_j},
\end{align}
which is the probability of departing from class $\mathcal{C}_i$ given the previous state is in $\mathcal{C}_i$. Then the following equations are straightforward
\begin{align}
p_{ii}:&=\Pb(X_1\in\mathcal{C}_i|X_0\in\mathcal{C}_i)=1-P_i, \nonumber \\
p_{ij}:&=\Pb(X_1\in\mathcal{C}_j|X_0\in\mathcal{C}_i)=\sum_{s\in\mathcal{C}_i}p_s p_{s,\mathcal{C}_j}, \quad \mbox{$j\neq i$}, \nonumber
\end{align}
where $\sum_{j}p_{ij}=1$. The following recursive formula holds
\begin{align}
p_{ik}^{(n)}:=\Pb(X_n\in\mathcal{C}_k|X_0\in\mathcal{C}_i)=\sum_{l=1}^{n}p_{ij}^{(l)}p_{jk}^{(n-l)}.
\end{align}
\tcr{Therefore, the state-dependent information freshness metric is}
\begin{align}
\Phi(n|x)=(n-\delta(n))H(P_T)+\sum_{k=1}^{n-\delta(n)}\Pb(T_i=0)^k(1-\Pb(T_i=0))^{n-\delta(n)-k} H(p_{ij}^{(k)})
\end{align}

\subsection{Proof of Proposition~\ref{eqn:prop1:1}}

The stationary distribution is obtained by solving balanced equation
\begin{align}
&\mu_{x,i}=\mu_{x,0}\prod_{j=0}^{i-1}(1-q_{xj}),\quad\forall x\in\mathcal{X},i\in\Zb_+, \\
&\mu_{x,0}=\sum_{y\in\mathcal{X}}\sum_{i=0}^\infty \mu_{y,i}q_{yi}p_{yx},\quad\forall x\in\mathcal{X}, \\
&\sum_{x\in\mathcal{X}}\sum_{i=0}^\infty \mu_{x,i}=1,
\end{align}
where $p_{xy}$ is the probability of transition from state $x$ to state $y$ under state change matrix $P_X$, $q_{xi}$ is the state change probability of state $(x,i)$. 

The entropy of recurrent Markov chain $(X_n,T_n)$ is
\begin{align}
&\sum_{x\in\mathcal{X}}\sum_{i=0}^\infty\mu_{x,i}h(1-q_{xi};\{q_{xi}p_{xy}\}_{y\in\mathcal{X}}) \nonumber\\
&=\sum_{x\in\mathcal{X}}\mu_{x,0}\sum_{i=0}^\infty\prod_{j=0}^{i-1}(1-q_{xj})h(1-q_{xi};\{q_{xi}p_{xy}\}_{y\in\mathcal{X}}).
\end{align}
For distribution $(1-q_{xi};\{q_{xi}p_{xy}\}_{y\in\mathcal{X}})$, the entropy is reduced to
\begin{align}
&h(1-q_{xi};\{q_{xi}p_{xy}\}_{y\in\mathcal{X}}) \nonumber \\
&=-(1-q_{xi})\log(1-q_{xi})-\sum_{y\in\mathcal{X}}q_{xi}p_{xy}\log(q_{xi}p_{xy}) \\
&=-(1-q_{xi})\log(1-q_{xi})-q_{xi}\log q_{xi} \sum_{y\in\mathcal{X}}p_{xy} \nonumber \\
&\qquad \qquad -q_{xi}\sum_{y\in\mathcal{X}}p_{xy}\log p_{xy} \\
&=h(q_{xi})+q_{xi}h(\mathbf{p_x}).
\end{align}

\subsection{Proof of Corollary~}
We observe that if $\{\mu_x\}_{x\in\mathcal{X}}$ is the stationary distribution of Markov chain with transition matrix $P_X$, then 
\begin{align}
\mu_{x,i}=\frac{\prod_{j=0}^{i-1}(1-q_{j})}{\sum_{k=1}^{\infty}\prod_{j=0}^{k-1}(1-q_{j})}\mu_x, \quad \forall x\in\mathcal{X},i\in\Zb_+,
\end{align}
is the stationary distribution for Markov chain $(X_n,T_n)$ with homogeneous $P_T$. 

\tcr{defer to appendix}
Since Markov chain $(X_n,T_n)$ is recurrent, we have
\begin{align}
\sum_{i=0}^{\infty}\prod_{j=0}^{i-1}(1-q_{j})q_{i}=1.
\end{align}
The balanced equation of Markov chain with transition matrix $P_X$ is
\begin{align}
&\mu_{x}=\sum_{y\in\mathcal{X}}\mu_{y}p_{yx}, \quad\forall x\in\mathcal{X}, \\
&\sum_{x\in\mathcal{X}} \mu_{x}=1,
\end{align}
while for $(X_n,T_n)$ we have
\begin{align}
&\mu_{x,i}=\mu_{x,0}\prod_{j=0}^{i-1}(1-q_{i}),\quad \forall x\in\mathcal{X}, i\in\Zb_+ \\
&\mu_{x,0}=\sum_{y\in\mathcal{X}}\sum_{i=0}^\infty\mu_{y,i}q_{i}p_{yx}, \quad\forall x\in\mathcal{X}, \\
&\sum_{x\in\mathcal{X}}\sum_{i=0}^\infty \mu_{x,i}=1.
\end{align}
One can verify the $\mu_{x,i}=\frac{\prod_{j=0}^{i-1}(1-q_{j})}{\sum_{k=1}^{\infty}\prod_{j=0}^{k-1}(1-q_{j})}\mu_x$ is the solution.
\tcr{defer to appendix}

Hence, we have
\begin{align}
&H(P_X,P_T) \nonumber\\ &=\frac{1}{\sum_{k=1}^{\infty}\prod_{j=0}^{k-1}(1-q_{j})}\sum_{x\in\mathcal{X}}\mu_x\sum_{i=0}^{\infty}\prod_{j=0}^{i-1}(1-q_{j})[h(q_i)+q_i h(\mathbf{p_x})] \\
&=\frac{\prod_{j=0}^{i-1}(1-q_{j})}{\sum_{k=1}^{\infty}\prod_{j=0}^{k-1}(1-q_{j})}h(q_i)+\frac{1}{\sum_{k=1}^{\infty}\prod_{j=0}^{k-1}(1-q_{j})}\sum_{x\in\mathcal{X}}\mu_x h(\mathbf{p_x}) \label{eqn:coro:1} \\
&\tcr{explain}=H(P_T)+\Pb(T_i=0)H(P_X)
\end{align}
where in Eqn.~(\ref{eqn:coro:1}) we use the identity $\sum_{i=0}^{\infty}\prod_{j=0}^{i-1}(1-q_{j})q_{i}=1$ for recurrent Markov chain $(X_n,T_n)$.
\fi

\bibliographystyle{IEEEtran}
\bibliography{ieeeabrv,AgeInfo}

\begin{thebibliography}{10}
\providecommand{\url}[1]{#1}
\csname url@samestyle\endcsname
\providecommand{\newblock}{\relax}
\providecommand{\bibinfo}[2]{#2}
\providecommand{\BIBentrySTDinterwordspacing}{\spaceskip=0pt\relax}
\providecommand{\BIBentryALTinterwordstretchfactor}{4}
\providecommand{\BIBentryALTinterwordspacing}{\spaceskip=\fontdimen2\font plus
\BIBentryALTinterwordstretchfactor\fontdimen3\font minus
  \fontdimen4\font\relax}
\providecommand{\BIBforeignlanguage}[2]{{%
\expandafter\ifx\csname l@#1\endcsname\relax
\typeout{** WARNING: IEEEtran.bst: No hyphenation pattern has been}%
\typeout{** loaded for the language `#1'. Using the pattern for}%
\typeout{** the default language instead.}%
\else
\language=\csname l@#1\endcsname
\fi
#2}}
\providecommand{\BIBdecl}{\relax}
\BIBdecl

\bibitem{infocom/KaulYG12}
S.~K. Kaul, R.~D. Yates, and M.~Gruteser, ``Real-time status: How often should
  one update?'' in \emph{{IEEE} {INFOCOM}}, Orlando, FL, USA, Mar. 2012, pp.
  2731--2735.

\bibitem{YatesK16}
R.~D. {Yates} and S.~K. {Kaul}, ``The age of information: Real-time status
  updating by multiple sources,'' \emph{IEEE Transactions on Information
  Theory}, vol.~65, no.~3, pp. 1807--1827, March 2019.

\bibitem{Pappas:2015:ICC}
N.~Pappas, J.~Gunnarsson, L.~Kratz, M.~Kountouris, and V.~Angelakis, ``Age of
  information of multiple sources with queue management,'' in \emph{IEEE
  International Conference on Communications (ICC)}, Jun. 2015, pp. 5935--5940.

\bibitem{isit/NajmN16}
E.~Najm and R.~Nasser, ``Age of information: The gamma awakening,'' in
  \emph{{IEEE} International Symposium on Information Theory (ISIT)},
  Barcelona, Spain, Jul. 2016, pp. 2574--2578.

\bibitem{tit/KamKNE16}
C.~Kam, S.~Kompella, G.~D. Nguyen, and A.~Ephremides, ``Effect of message
  transmission path diversity on status age,'' \emph{{IEEE} Trans. Inf.
  Theory}, vol.~62, no.~3, pp. 1360--1374, Mar. 2016.

\bibitem{Sun:ISIT:2017}
Y.~Sun, Y.~Polyanskiy, and E.~Uysal-Biyikoglu, ``Remote estimation of the
  {Wiener} process over a channel with random delay,'' in \emph{IEEE
  International Symposium on Information Theory (ISIT)}, Jun. 2017, pp.
  321--325.

\bibitem{Jiang:2019:INFOCOM}
Z.~{Jiang}, S.~{Zhou}, Z.~{Niu}, and C.~{Yu}, ``A unified sampling and
  scheduling approach for status update in multiaccess wireless networks,'' in
  \emph{IEEE INFOCOM}, April 2019, pp. 208--216.

\bibitem{isit/BedewySS16}
A.~M. Bedewy, Y.~Sun, and N.~B. Shroff, ``Optimizing data freshness,
  throughput, and delay in multi-server information-update systems,'' in
  \emph{{IEEE} International Symposium on Information Theory (ISIT)},
  Barcelona, Spain, Jul. 2016, pp. 2569--2573.

\bibitem{infocom/SunUYKS16}
Y.~Sun, E.~Uysal{-}Biyikoglu, R.~D. Yates, C.~E. Koksal, and N.~B. Shroff,
  ``Update or wait: How to keep your data fresh,'' in \emph{{IEEE INFOCOM}},
  San Francisco, CA, USA, Apr. 2016, pp. 1--9.

\bibitem{He:2018:TIT}
Q.~He, D.~Yuan, and A.~Ephremides, ``Optimal link scheduling for age
  minimization in wireless systems,'' \emph{IEEE Transactions on Information
  Theory}, vol.~64, no.~7, pp. 5381--5394, July 2018.

\bibitem{Kadota:2018:INFOCOM}
I.~Kadota, A.~Sinha, and E.~Modiano, ``Optimizing age of information in
  wireless networks with throughput constraints,'' in \emph{IEEE INFOCOM}, Apr.
  2018.

\bibitem{Wang:JCN:2019}
B.~Wang, S.~Feng, and J.~Yang, ``When to preempt? age of information
  minimization under link capacity constraint,'' \emph{Journal on
  Communications and Networking}, vol.~21, no.~3, pp. 220--232, Jun. 2019.

\bibitem{Zhou:TWC:2019}
B.~{Zhou} and W.~{Saad}, ``Minimum age of information in the internet of things
  with non-uniform status packet sizes,'' \emph{IEEE Transactions on Wireless
  Communications}, pp. 1--1, 2019.

\bibitem{Yang:2017:AoI}
X.~Wu, J.~Yang, and J.~Wu, ``Optimal status update for age of information
  minimization with an energy harvesting source,'' \emph{IEEE Transactions on
  Green Communications and Networking}, vol.~2, no.~1, pp. 193--204, March
  2018.

\bibitem{Yang:2018:IT}
A.~{Arafa}, J.~{Yang}, S.~{Ulukus}, and H.~V. {Poor}, ``{Age-Minimal
  Transmission for Energy Harvesting Sensors with Finite Batteries: Online
  Policies},'' \emph{IEEE Trans. on Information Theory}, vol.~66, no.~1, pp.
  534--556, Jan. 2020.

\bibitem{Feng:TAC:arXiv}
S.~Feng and J.~Yang, ``Age of information minimization for an energy harvesting
  source with updating erasures: With and without feedback,'' \emph{CoRR}, vol.
  abs/1808.05141, 2018.

\bibitem{Yates:2017:ISIT}
R.~D. Yates, E.~Najm, E.~Soljanin, and J.~Zhong, ``Timely updates over an
  erasure channel,'' in \emph{2017 IEEE International Symposium on Information
  Theory (ISIT)}, Jun. 2017, pp. 316--320.

\bibitem{Mayekar2018}
P.~Mayekar, P.~Parag, and H.~Tyagi, ``Optimal lossless source codes for timely
  updates,'' \emph{IEEE International Symposium on Information Theory (ISIT)},
  pp. 1246--1250, 2018.

\bibitem{Emina:2018}
J.~{Zhong}, R.~D. {Yates}, and E.~{Soljanin}, ``{Timely Lossless Source Coding
  for Randomly Arriving Symbols},'' \emph{ArXiv e-prints}, Oct. 2018.

\bibitem{Feng:Globecom:2019}
S.~Feng and J.~Yang, ``Adaptive coding for information freshness in a two-user
  broadcast erasure channel,'' in \emph{IEEE Global Communications Conference
  (Globecom)}, Hawaii, USA, Dec. 2019.

\bibitem{Baknina:2018:CISS}
A.~{Baknina} and S.~{Ulukus}, ``Coded status updates in an energy harvesting
  erasure channel,'' in \emph{Conference on Information Sciences and Systems
  (CISS)}, Mar. 2018.

\bibitem{Zhong:ISIT:2018}
J.~Zhong, R.~D. Yates, and E.~Soljanin, ``Two freshness metrics for local cache
  refresh,'' in \emph{IEEE International Symposium on Information Theory
  (ISIT)}, June 2018, pp. 1924--1928.

\bibitem{maatouk2019age}
A.~Maatouk, S.~Kriouile, M.~Assaad, and A.~Ephremides, ``The age of incorrect
  information: A new performance metric for status updates,'' \emph{arXiv
  preprint arXiv:1907.06604}, 2019.

\bibitem{Sun:2018:SPAWC}
Y.~{Sun} and B.~{Cyr}, ``Information aging through queues: A mutual information
  perspective,'' in \emph{2018 IEEE 19th International Workshop on Signal
  Processing Advances in Wireless Communications (SPAWC)}, June 2018, pp. 1--5.

\bibitem{singh2019optimal}
R.~Singh, G.~K. Kamath, and P.~R. Kumar, ``Optimal information updating based
  on value of information,'' 2019.

\bibitem{Kosta:2017:nonlinear}
A.~Kosta, N.~Pappas, A.~Ephremides, and V.~Angelakis, ``Age and value of
  information: Non-linear age case,'' in \emph{IEEE International Symposium on
  Information Theory (ISIT)}, Aachen, Germany, Jun. 2017, pp. 326--330.

\end{thebibliography}

\end{document}